\documentclass[12pt, preprint]{aastex}
\usepackage{natbib}

\begin{document}

\title{The Stability and  Dynamics of Planets in Tight Binary Systems}

\author{Lamya A. Saleh \ and Frederic A. Rasio}

\affil{Department of Physics \& Astronomy , Northwestern University, Evanston, IL 60208}
\email{l-saleh@northwestern.edu}

\begin{abstract}
 Planets have been observed in tight binary systems with separations less than $20$ AU. A likely formation scenario for such systems involves a dynamical capture, after which high relative inclinations are likely and may lead to Kozai oscillations. 
  We numerically investigate the fate of an initially coplanar double-planet system in a class of binaries with separation ranging between $12 - 20 $ AU.  Dynamical integrations of representative four-body systems are performed, each including a hot Jupiter and a second planet on a wider orbit. We find that,  although such systems can remain stable at low relative inclinations ($\lesssim 40^\circ$), high relative inclinations are likely to lead to instabilities. This can be avoided if the planets are placed in a  \emph{Kozai-stable zone} within which mutual gravitational perturbations can suppress the Kozai mechanism.  
  We investigate the possibility of inducing Kozai oscillations in the inner orbit by a weak coupling mechanism between the planets in which the coplanarity is broken due to a differential nodal precession.   Propagating perturbations from the stellar companion through a planetary system in this manner can have dramatic effects on the dynamical evolution of planetary systems, especially in tight binaries and can offer a reasonable explanation for eccentricity trends among planets observed in binary systems.  We find that inducing such oscillations into the orbit of a hot Jupiter is more likely in tight binaries and an upper limit can be set on the binary separation above which these oscillations are not observed. 
   \end{abstract}

\section{Introduction}
 
  Among all extrasolar planets discovered to date, $20 \%$  are in multiple stellar systems \citep{des07,egg07b}. 
Binary stellar systems with circumstellar and
circumbinary debris disks have been observed with an
 overall occurrence rate higher than that for debris disks around single stars \citep{tri06}. 
 Numerical simulations suggest that protoplanetary disks embedded in binary star systems should be able to form both terrestrial and giant planets \citep{bar02,qui02,bos06}. 

  Most binary and multiple stellar systems found to be harboring planets are wide, with separations larger than $100$ AU \citep{egg04,mug04}. However, several systems with separations as low as  $\sim 20$ AU have been shown to harbor giant  planets. These are the binary systems HD196885, $\gamma$ Cephei, and Gliese 86, and the higher order system HD41004
  \citep{cor08,hat03,els01,zuc04}.  There are two important things to note about these systems.
  First, since radial-velocity surveys  have always been in biased against close binaries \citep{egg07b},
this sample probably underrepresents the frequency of planets in such binaries.  In particular,  the lower limit for the separation of binaries that can harbor planets may be smaller than $\sim 20$ AU.  Second, such systems constitute a unique data set which can allow us to test theoretical models for  
planet formation and evolution. This is because the presence of a stellar 
companion at such proximity can greatly influence these processes.
 A stellar companion as far as $\sim 50$ AU can weaken the chances for the formation of a giant planet by stirring, heating or truncating  protoplanetary disks \citep{jan07,bos06,may05,kle01,nel00a,nel00b}.  Therefore, the fact that systems as tight as $20$ AU are harboring giant planets is puzzling and  does seem to require further investigation.  The most likely formation scenario for such systems may involve dynamical encounters  in dense stellar systems where the binary companion is captured or moved inward from a wider orbit after the planet formation has taken place.  
 A stellar companion can also affect the dynamical evolution of
a planetary system, and therefore its long term survival,  through secular perturbations.   In the case of a large relative inclination 
between the planetary orbit and that of the stellar companion, Kozai oscillations can take place where angular momentum exchange between the orbits leads to large-amplitude synchronous oscillations in the eccentricity and 
inclination of the planet \citep{koz62,holm97,inn97,for00,tak05}. This has the potential to disrupt the system if it results
in close encounters, orbital crossings, or strong planet-planet interactions for systems with more than one planet.  
 
A renewed interest in planetary dynamics in general and in the stability of planets in binary systems particularly has been triggered by the discovery of planets in binary and multiple stellar systems \citep{dvo03,bar04,ray05,ray06,riv07}.  
  Looking into the dynamical history of planetary systems is key to understanding the processes involved
in their formation and subsequent evolution.  In fact, this problem has received special interest from dynamicists even 
before extrasolar planets were discovered, with many studies applying general results of the three body problem to the special case of a binary stellar system harboring a planet \citep[e.g,][hereafter HW99]{dvo93,inn97,hol99}.
By performing numerical simulations on planetary orbits in binary systems, HW99 derived empirical expressions for the maximum semi-major axis of 
the planet as a function of both binary mass fraction and eccentricity, above
 which the system would become unstable. Their study considered the stability of binary systems 
harboring planets in the coplanar case only. 
 Numerical simulations by \citet{dav03} of Earth-like planets in binary
 systems with different initial configurations led to the estimate that 50\% of binary systems 
allow an Earth like planet to remain stable for 4.6 Gyr. 
 The stability of Earth-like planets in binaries was also studied by \citet{fat06}.  With numerical simulations they calculated the survival time of an Earth-like planet orbiting a 
Sun-like star in the presence of a stellar companion. They explored a range of 
binary configurations including relative inclinations between the binary orbit and the
orbit of the planet. 
 \citet{hag06} performed numerical simulations to model the binary system $\gamma$ Cephei. This is a tight binary with separation $18.5$ AU and an eccentricity of 
$0.36$.  A jovian planet with a minimum mass of $1.7 M_{J}$ was shown to be orbiting the primary of this system at $2.1$ AU \citep{hat03}.  
  The question of the presence of an Earth-like planet within the habitable zone (HZ) of the primary was addressed and a range of semi-major axes and eccentricities of the binary as well as orbital inclination of the Jupiter-like planet was explored.   This study concludes that, within timescales in the range  $10-100$ Myr,  such a planet can exist on a stable orbit but not within the HZ of the primary.

  In this study, we investigate the stability of double-planet systems in tight binaries. 
  The tightest stellar system in which a planet was ever claimed is HD188753.  This is a hierarchical triple star system in which the primary is  a Sun-like star with mass $1.06$ $M_\odot$. The binary  companion is a double-star system with total mass $1.63$ $M_\odot$ orbiting the primary at separation  $\sim{12.3}$ AU  and eccentricity 0.5 \citep{gri77}.  A hot Jupiter of mass $1.14$ $M_{J}$ was claimed to be orbiting the primary star with a period of 3.35 d  \citep{kon05}. Because of the proximity of the companion, the presence of a planet in this system is suggestive of a dynamical history in which the binary companion was either moved inward to a tighter orbit or captured  after the claimed planet had already formed and reached its current position \citep{por05,pfa05,bos06,jan07}. 
Although the observation of this planet was refuted by  \citet{egg07a},  we find this system to be a good representative of the lower limit (in separation) of a class of tight binary systems harboring giant planets.  We test in this study the capacity of such systems to harbor two planets including a hot Jupiter.
 This class of binaries with separation $\lesssim 20$ AU is of special interest since it is likely that such systems may have had active dynamical histories and are expected to induce large perturbative forces to planetary systems with dramatic results.  Studying such systems  in the presence of more than one planet, can uncover the role played by binary companions in shaping planetary systems. 
This  is especially important since multi-planet systems seem to be common in nature \citep{fis03}. 
The dynamics of double-planet systems in wide binaries was studied analytically by \citet{tak08}, where three different dynamical classes were identified as possible outcomes: (1) Decoupled systems in which planetary orbits experience independent Kozai cycles due to their weak mutual interaction compared to perturbations from the companion; (2) Weakly coupled systems in which mutual gravitational interactions become more dominant and relative nodal precession can cause mutual inclinations to grow and induce Kozai oscillations in the inner orbit; (3) Dynamically rigid systems in which the orbital elements of both planets oscillate in concert.
These possible outcomes depend on the initial conditions and therefore the relative strength of the different perturbative forces. 
Here we expand on this study by looking into systems with closer companions and test the possibility of inducing eccentricity oscillations in a hot Jupiter orbit through a weakly coupled system.   

Since planets with orbital radii $a \lesssim 0.1$ AU are expected to be tidally circularized, the recent observations of a few hot Jupiters on considerably eccentric orbits (such as XO-3, HAT-P-2, and HD185269 \citep{joh08,loe08,joh06}) does raise questions.  Why are the orbits of these planets not circularized when the orbits of most hot Jupiters are?  \citet{mat08} show that close-in planets on eccentric orbits can be explained by constraining their tidal $Q$ factors under the assumption that these orbits are in the process of being circularized.   They also explore the possibility that these eccentricities are induced by a perturber on a wider orbit for the system GJ436 and show that such a planet, if present, would cause a radial-velocity amplitude above the current detection limit, and therefore, would have been observed.  

  Although such eccentricities can, in principle, be explained by the presence of a companion, whether stellar or sub-stellar, one has to keep in mind that a planet on a close-in orbit experiences strong GR forces. Therefore, a stellar companion as close as $12 - 20$ AU cannot induce such eccentricities directly into a hot Jupiter orbit.  
We test here the possibility of inducing such eccentricities to a hot Jupiter through a second planet in the presence of a binary companion. 
We  also investigate the dynamics and stability criteria of a double-planet system orbiting the primary of a tight binary system with separation in the range $ a_{b} = 12 - 20$ AU.  
All systems we investigate in this study contain a hot Jupiter of mass $1 M_{J}$ and separation $a_{1} = 0.05$ AU.  The stability region for a second planet, placed on a wider orbit, that is initially circular and coplanar  with a variable semi-major axis $a_{2}$, was explored for two different binary configurations.   In the first, the binary companion orbits at  $ a_{b} = 12$ AU  with an eccentricity $ e_{b} = 0.5$, and in the second, $ a_{b} = 20$ AU and $ e_{b} = 0$.  For the tightest binary separation  ($ a_{b} = 12$ AU), two different planetary masses $m_{2}=M_{J}$  and $m_{2} =M_{\oplus}$ were investigated. 
In this class of systems, planets experience mutual gravitational perturbations in addition to perturbations from the stellar companion. A hot Jupiter orbit is strongly dominated by GR forces. Because of the tightness of these systems, the interplay between these different forces is expected to have a dramatic effect on the planetary system. 

Throughout this paper, the subscripts 0, 1, 2, and b will be used to refer to the primary star, the inner planet, the outer planet, and the binary companion, respectively. 
 We use a semi-analytical approach to predict the stability region of the
 outer planet $m_{2}$. 
Our numerical simulations incorporate   
 a wide range of parameters including mutual inclinations of the orbits and perturbative forces to the Keplerian motion of each planet including planet-planet interactions and GR effects.    
Such numerical simulations can reveal to us the likelihood of a compact stellar system to maintain multiple planets and can be a useful guide for future observational efforts such as HARPS and the Kepler Mission.

 An overview of secular perturbations affecting this class of systems is presented in \S 2 . In \S 3 we discuss the analytical background for our predicted stability region for the outer planet in a stellar system resembling HD188753.  The numerical techniques used in this study are presented in \S 4, results and discussion in \S 5, and finally,  \S 6 summarizes our conclusions and suggestions for future work.\\

\section{Secular Perturbations in Hierarchical Triple Systems}

     Hierarchical triple systems can be studied using classical Hamiltonian perturbation techniques, which show that the secular evolution of the orbits takes the form of eccentricity oscillations coupled  with precession of the longitudes of pericenter  \citep{bro61,koz62,har68,ras97,for00}.
 Such perturbations are highly dependent on the initial relative inclination between the two orbits. The classical planetary perturbation theory  \citep{bro61} 
 applies to the case of low relative inclination orbits and low eccentricities for planets around a central star  and is applicable to all orders in the ratio of the semi-major axes  $\alpha$, where $\alpha = a_{1}/ a_{2}$,  where $ a_{1}$ and $a_{2}$ represent the semi-major axes of  the inner and the outer orbits respectively \citep[e.g,][]{ras94,ras95}.  
 The secular evolution of a triple system with large relative inclination was derived by \citet{koz62} for the limit of small $\alpha$ using the quadrupole approximation of secular perturbation theory. 
 In the following we discuss and calculate precession rates associated with these perturbations in addition to GR forces as they apply  to the type of systems investigated in this study (see \S 1). 

   \subsection{The Low Inclination regime}
Mutual gravitational perturbations among planets  can lead to a precession of the periastra and eccentricity oscillations of both planets.  In the case of low eccentricities and low relative inclinations the periods of these oscillations, can be calculated using the formulation of the Laplace-Lagrange secular theory in which the disturbing potentials of the planets are simplified by eliminating short-period terms (dependent on the mean longitudes).  Following the classical planetary perturbation theory ( \citet{bro61}, see also Murray \& Dermott 1999) for a two-planet system, the disturbing functions for the two planets, to second order in eccentricity, are
\begin{equation}
    {R}_{1} = {n}_{1} {a}_{1}^{2} \left[\frac{1}{2}      {A}_{11} e_{1}^{2} +      {A}_{12} e_{1} e_{2} \cos(\omega_{1} - \omega_{2}) \right]
\end{equation}
and
\begin{equation}
    {R}_{2} = {n}_{2} {a}_{2}^{2} \left[\frac{1}{2}      {A}_{22} e_{2}^{2} +      {A}_{21} e_{1} e_{2} \cos(\omega_{1} - \omega_{2}) \right],
\end{equation}
where $n$, $a$, and $\omega$, refer to the mean motions, semi-major axes, and  longitudes of pericenter of the planets, respectively.  The matrix elements 
\begin{equation}
      {A}_{11} = \frac{1}{4} n_{1} \frac{m_{2}}{m_{0} + m_{1}}\alpha^{2}b_{3/2}^{(1)}(\alpha),
 \end{equation}
 
\begin{equation}
      {A}_{22} = \frac{1}{4} n_{2} \frac{m_{1}}{m_{0} + m_{2}}\alpha b_{3/2}^{(1)}(\alpha),
 \end{equation}
 and
\begin{equation}
      A_{0}=\frac{{A}_{11}}{{A}_{12}}=  \frac{ {A}_{21}}{ {A}_{22}} \approx -  \frac{5}{4} \alpha (1 - \frac{1}{8}\alpha^{2}) ,
 \end{equation}
are functions of the planetary masses and semi-major axes, where $\alpha = a_{1}/a_{2}$, and
\begin{equation}
   b_{3/2}^{(1)}(\alpha) = \frac{1}{\pi} \int_{0}^{2\pi}\frac{\cos\psi \hspace{0.05in} d\psi}{(1-2\alpha\cos\psi +\alpha^{2})^{3/2}}
 \end{equation}

 The eigenvalues of the matrix $ {A}$ represent the characteristic frequencies of the system with which the planetary orbital elements will evolve. These eigenfrequencies are given by
\begin{equation}
  g^{+} = \frac{1}{2} \left[ \;(A_{11} + {A}_{22})  + \sqrt{(A_{11} - A_{22})^{2} + 4A_{0}^{2}A_{11}A_{22} } \; \right],
 \end{equation}
 and
 \begin{equation}
  g^{-} = \frac{1}{2} \left[ \;(A_{11} - {A}_{22})  + \sqrt{(A_{11} - A_{22})^{2} + 4A_{0}^{2}A_{11}A_{22} } \; \right].
 \end{equation}
 Pericenter precession rates are ,in general, non-linear combinations of these eigenvalues and can be estimated for different planetary systems \citep{tak08}.  The periods of oscillations can then be calculated from these rates.  In this study, we use the period of orbital eccentricity oscillations given by $ P_{e} = \frac{2\pi}{| g^{+} - g^{-}|}$.

 \subsection{The High Inclination regime}
   \citet{koz62}  showed that in a hierarchical system where the ratio of semi-major axes, $\alpha$, is sufficiently small and a large relative inclination ($\gtrsim 40^{\circ}$) between the orbits exists,  exchange of angular momentum between the orbits results in simple periodic oscillations in both the inner eccentricity and the mutual inclination of the orbits \emph{i} such that they are coupled by the integral of motion $(1-e^{2})\cos^{2}{i}$.
Under these conditions the amplitude of eccentricity oscillations depends only on the relative inclination $i$  \citep{inn97} and is given at quadrupole order by,
\begin{equation}
  e_{max} = \sqrt{ 1 - \frac{5}{3} \cos^{2}{i} }
\end{equation}

 The period of oscillations is a function of  $\alpha$, the masses, and the eccentricity of the outer orbit \citep{kis98}
\begin{equation}
 P_{Kozai} \simeq P_2 \left( \frac{m_0+m_2}{m_b}\right)\left(\frac{a_b}{a_2}\right)^{3}\left(1-e_b^{2} \right)^{3/2},
\end{equation}
 where $P_2$ is the orbital period of the planet.

 \subsection{GR Forces}

      For close-in planets, general relativistic effects become dominant and can cause the periastron of the planet's orbit to precess on very short  timescales  \citep{egg01},
 \begin{equation}
 P_{GR} \simeq 3011\hspace{0.05in} yr \hspace{0.08in} \left(\frac{a}{0.05AU}\right)^{5/2} \left(\frac{m_0}{1.0M_\odot}\right)^{3/2},
\end{equation}
where $a$ is the semi-major axis of the planet and $m_0$ is the mass of the central star.
  This precession can lead to the suppression of Kozai oscillations by averaging out the torques responsible for it if they act on shorter timescales.

  \subsection{Comparison of Timescales}
  
 We compare in Figure 1 the precession rates due to all three perturbations acting on both planets in the case of a double-planet system in a binary with separation $a_{b} = 12$ AU and eccentricity $e_{b} =0.5$.  The planets, with masses $m_{1}=m_{2}=M_{J}$, are placed initially on circular and coplanar orbits with $a_{1}= 0.05$ AU. 
  \begin{figure}
\epsscale{0.85}
\plotone{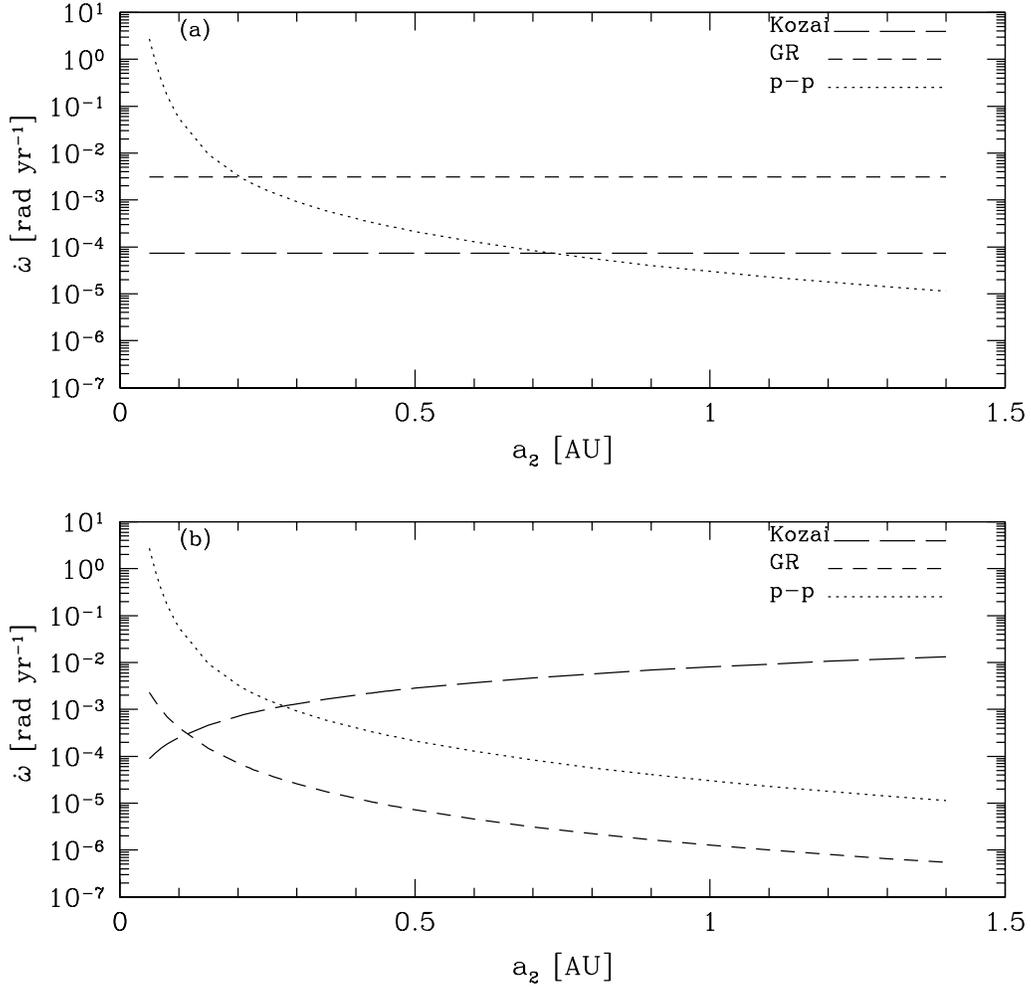}
\caption{Secular precession rates, calculated from the periods such that $\dot{\omega} = 2 \pi / P$, for all perturbations acting on the inner planet (a), and on the outer planet (b) in a double-planet system with   $m_{1}=m_{2}=M_{J}$, and $a_{1}= 0.05$ AU. The binary parameters are  $a_{b} = 12$ AU, $e_{b} =0.5$.  Perturbations include planet-planet interactions (p-p),  GR effects, and the Kozai secular perturbation.}
\end{figure}

  The dynamics of the outer planet are expected to be dominated by the Kozai effect as long as the semi-major axis of its orbit is larger than $ a_{2} \sim 0.3$ AU.  Eccentricity oscillations with large amplitudes can cause this planet to experience close encounters either with the companion or with the hot Jupiter, leading to instabilities. 
Below $ a_{2} \approx 0.3$ AU,  mutual gravitational interactions among the planets dominate, and Kozai oscillations are gradually suppressed. Therefore, a \emph{Kozai-stable zone} exists below this point, within which this second planet can maintain stability even when relative initial inclinations are larger than the Kozai limit $\emph{i}_{init} \approx 40^\circ$. 
The inner orbit at $a_{1}=0.05$ AU is strongly affected by GR forces such that   $\dot{\omega}_{GR} >\dot{\omega}_{Koz}$ and Kozai oscillations are suppressed  in this planet.   On the other hand, mutual planetary interactions can dominate the dynamics of the system as the outer planetary orbit gets tighter than $a_{2} \lesssim 0.2$ AU and  periastron precession rates associated with this force $\dot{\omega}_{pp}$ exceed those associated with GR forces $\dot{\omega}_{GR}$. This should be revealed in the form of mild eccentricity oscillations in both orbits.   

  We look into the possibility of inducing Kozai oscillations in the inner orbit by propagating perturbations from the companion through the outer planet in the system.   Kozai oscillations can be induced in the inner orbit if a relative nodal precession occurs,  splitting the planes of the planetary orbits and producing  relative inclinations larger than the Kozai critical angle.
For these oscillations to occur two conditions must be satisfied: (1)   The outer planet's orbit must experience nodal precession causing periodic growth in the mutual inclination angle between the planets. The outer planet can experience nodal precession even if it is not experiencing Kozai oscillations since this precession is a natural consequence of secular interactions with the stellar companion \citep{tak08};  (2) Precession rates associated with the induced Kozai oscillation of the inner orbit must exceed GR precession rates.   Whether both conditions can be satisfied simultaneously depends on the choice of initial conditions. In our systems the inner planetary orbit is fixed at $0.05$ AU. The outer planet must be placed in a region where it is close enough for the induced Kozai effect to compete with GR forces.
We show in Figure 2 the timescales for precession in the inner orbit caused by GR forces (eq.[11]) and 
 \begin{figure}
\epsscale{0.85}
\plotone{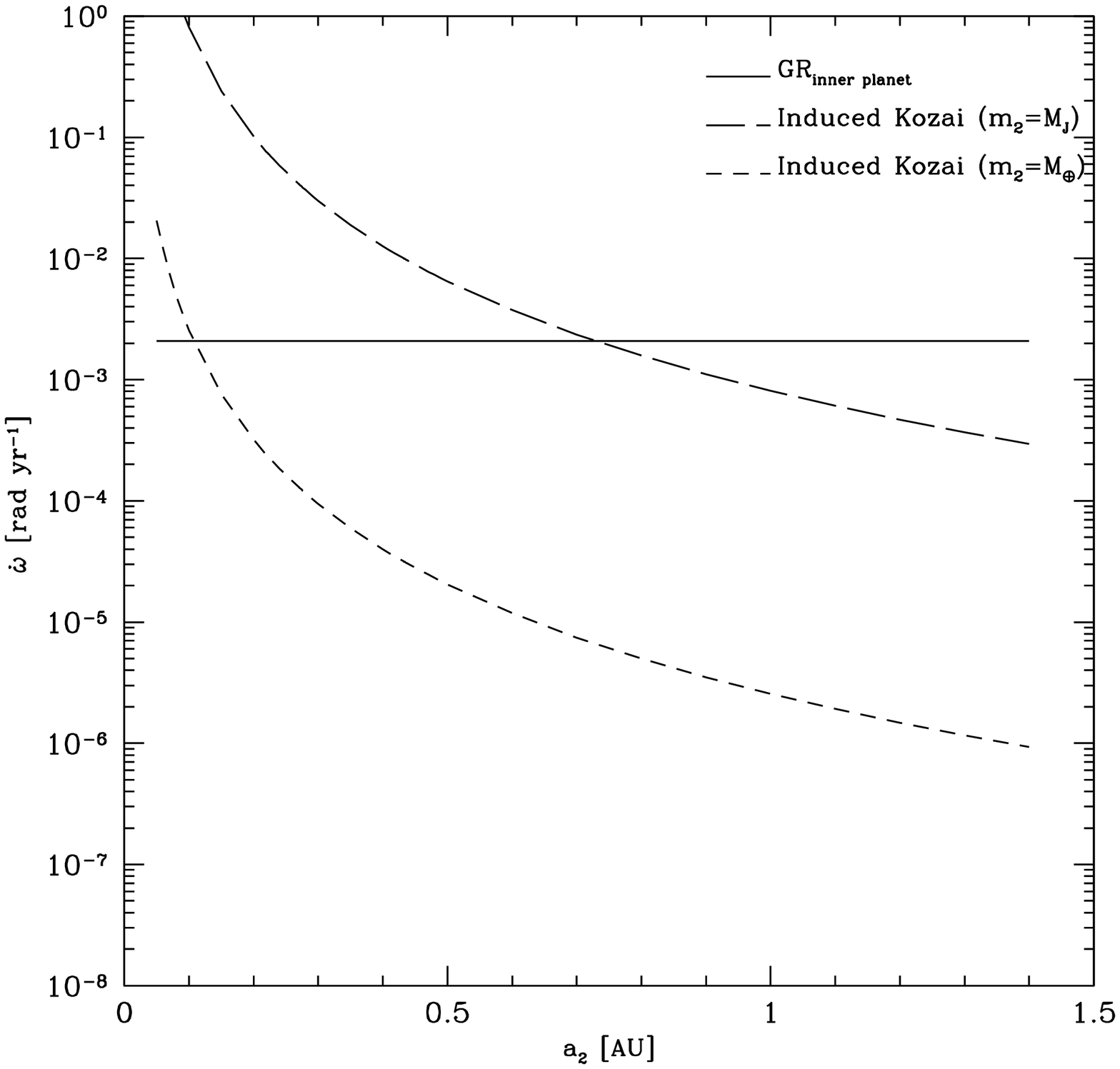}
\caption{The rate of precession in the inner orbit due to the GR effect compared to that associated with the mutual planetary Kozai effect for two different perturber masses   $m_{2}=M_{J}$ and $m_{2}=M_{\oplus}$.  System parameters are the same as in Figure 1.  The shown forces are: the GR effect (solid), the mutual planetary Kozai effect for $m_{2}=M_{J}$ (long-dashed), and for  $m_{2}=M_{\oplus}$ (short-dashed).  For equal mass planets, these two timescales become comparable at $a_{2} \approx 0.8$ AU beyond which the induced Kozai cycles disappear. In the case of an Earth-mass planet in the outer orbit, mutual planetary Kozai perturbations become too weak to overcome GR forces.}
\end{figure}
those caused by the mutual planetary Kozai effect calculated from equation (10) by replacing  $m_{b}$ with  $m_{2}$ as the perturber and replacing $m_{2}$ with  $m_{1}$. Two different perturber masses are tested  $m_{2}=M_{J}$ and $m_{2}=M_{\oplus}$. 
 It shows that while an Earth-mass planet cannot compete with the strength of GR forces acting on such a tight inner orbit, a Jupiter-mass planet can produce this effect when  $a_{2}\lesssim 0.8$ AU.

\section{Analytical Predictions for Stability of a Double-Planet System in a tight Binary}

  We investigate the stability criteria of a double-planet system in a binary with separation $a_{b} \leq 20$ AU. The inner planet is a hot Jupiter with semi-major axis $a_{1}=0.05$ AU and mass $m_{1}=M_{J}$.  An approximate stability region for the outer planet is derived analytically within the two dimensional space [$i_{init}, a_{2}$] where $a_{2}$ is its semi-major axis and  $i_{init}$ is the initial relative inclination between the common planetary orbit and that of the binary.  This stability region is derived analytically for two binary separations:  $a_{b} =12$ AU and  $a_{b} = 20$ AU. In the case  $a_{b} =12$ AU, the stability region is  compared for two different planetary masses $m_{2} = M_{J}$ and $m_{2} =  M_{\oplus}$ in the outer orbit.
  For a given value of $i_{init}$, a planetary orbit is expected to be stable if  $a_{2}$ lies between  two limits, $a_{2,\rm min}$ and $a_{2,\rm max}$.   The lower limit, $a_{2,\rm min}$, is defined as the minimum semi-major axis below which strong perturbations with the inner orbit can lead to instabilities,  while the upper limit, $a_{2,\rm max}$,  is the maximum 
semi-major axis of the planet beyond which the planet enters a region of instability due to the presence of the stellar companion.

\subsection{Upper Limit}

  The upper limit for the semi-major axis of a planet in a binary system was determined numerically by 
 HW99 for  initially circular planetary orbits which are coplanar with the binary orbit.  This was done for a range of binary mass fractions $\mu_{b}=m_{b}/(m_{0}+m_{b})$ (where $m_{0}$ is the mass of the primary and $m_{b}$ is the mass of the perturber) and eccentricities $e_{b}$ of the binary. For  each set of values [$\mu_{b}$ , $e_{b}$], the binary system was numerically integrated with test particles representing the planet, and a critical semi-major axis $a_{c}$ was determined beyond which a planetary orbit may become unstable. An analytical expression  for $a_{c}$ was derived from these numerical results such that, 
\begin{equation}
a_{c} \approx (0.464 - 0.380\mu_{b} - 0.631e_{b} + 0.586e_{b}\mu_{b} + 0.150e_{b}^{2} - 0.198e_{b}^{2}\mu_{b})a_{b}
\end{equation} 
where $a_{b}$ is the semi-major axis of the binary.\\ 
Applying equation (12) to the outer planet in the  binary systems we consider gives a value of  $a_{c} \approx 1.23$ AU for the stellar system HD188753, and $a_{c} \approx 5.48$ AU for the equal mass binary with separation $20$ AU, both in the coplanar case.  
When large relative inclinations ($ i \geq 40^{o}$) are present, we estimate for an elliptical orbit the upper limit on the planet's semi-major axis using,
\begin{equation}
a_{2,{\rm max}} \approx \frac{a_{c} }{ ( 1 + e_{max})}
\end{equation} 
where $e_{max}$ is the maximum eccentricity gained during Kozai oscillations (eq.[1]), and assuming the value of $a_{c}$ calculated from equation (12) is the distance  from the primary at apastron.

\subsection{Lower Limit}
 The stability of three-body systems was discussed analytically in the literature in more than one context. Some aimed at the stability limits for a system composed of a central star orbited by two planets  \citep[e.g,][hereafter G93]{gla93}, while others were mainly focused on triple stellar systems  \citep[e.g,][hereafter MA01]{mar01}. \\
     G93 provides stability criteria for a system of one star with two planets. He derives a minimum separation between the two planets "$\delta$" below which the system cannot maintain Hill stability.  A system is considered \emph{Hill stable} if the planets cannot experience close encounters 
  at any time.
 For systems in which the central star $m_{0}$ is much more massive than both planets, $m_{1}$ and $m_{2}$
 ($\mu_{1} = {m_{1}}/{m_{0}} << 1$  and $ \mu_{2} = {m_{2}}/{m_{0}} << 1$), and for initially circular and nearly coplanar orbits,
 the minimum separation $\delta$, expressed in units such that the semi-major axis of the inner planet is unity, is given to lowest order in the masses as 
 \begin{equation}
 \delta \simeq 2.40(\mu_{1}+ \mu_{2})^{1/3}.
\end{equation}
For equal mass planets this reduces to  $\delta \simeq 3 \mu^{1/3}$.
  For non-circular orbits, but small eccentricities ($e \leq  \mu^{1/3}$) and equal-mass planets, this becomes, 
  \begin{equation}
 \delta \simeq \sqrt{ \frac{8}{3}(e_{1}^2 + e_{2}^2)+ 9\mu^{2/3}}.
\end{equation}
This reduces to the previous equal-mass case for zero eccentricities. \\
       G93 also derived the Hill stability criterion for the case of equal mass-planets and equal and arbitrarily \emph{large} eccentricities ($e_{1} = e_{2} = e$), giving
 \begin{equation}
 \delta \simeq \left( \sqrt{ \frac{(3 + e^2)}{2(1-e^2)} - \frac{1}{2} \sqrt{\frac{(9 -e^2)}{(1 - e^2)}}} +  \frac{1}{2}  \sqrt{ \frac{(9 -e^2)}{(1 - e^2)}} -  \frac{1}{2}  \right)^2  - 1 +  \mathcal{O}  (\mu^{1/3}). 
\end{equation}
  This expression is valid only for eccentricities larger than $ e \simeq  \mu^{1/3}$, in which case the mass terms become negligible compared to the large eccentricity terms. Therefore, at large eccentricities $\delta$ becomes independent of planetary masses. 

 One can apply this stability criterion to the two planets in our systems.  Eccentric orbits are expected  if the binary orbit is highly inclined relative to the common planetary plane. 
  In this case, $\delta$ is calculated from equation (16) by substituting $e_{max}$ from equation (1)  for the eccentricity. This sets a lower limit on the minimum separation allowed between the planets, since in some cases, the outer planetary orbit may experience Kozai oscillations while the inner planet is shielded from this effect due to stronger perturbations (\S 2.4).  To calculate $\delta$ for systems with relative inclinations less than $40^{\circ}$,  assuming $e_{1} = e_{2} \approx $ zero, we use equation (15) in the case of equal mass planets ($m_{1} = m_{2} =  M_{J}$) and equation (14) in the case of a second planet of Earth-mass ($m_{2} = M_{\oplus}$).
 Once $\delta $ is estimated, the minimum semi-major axis allowed for the second planet can be calculated as $a_{2,\rm min} =  a_{1}{(1 + \delta)}$, 
where  $a_{1} $ is the semi-major axis of the inner planet.
   The difference in the value of $a_{2, \rm min}$ obtained when applying the G93 criterion to the two different  values of $m_{2}$ used in this study, is $\sim 0.01$ AU. This is small enough that in practice here we ignore it when considering our predicted stability region. 

   The stability of a hierarchical triple configuration was discussed in
   MA01,  who describe hierarchical coordinates for an inner binary with masses $m_{0}$ and $m_{1}$, and orbit described by $e_{in}$ and $a_{in}$ (the inner eccentricity and semi-major axis). A more distant object $m_{2}$ orbits around the center of mass of the inner binary with eccentricity $e_{out}$ and periastron separation $R_{per}$. 
  They derive a critical value for ${R_{per}}/{a_{in}}$ below which the triple configuration can become unstable and may experience the escape of one object, 
\begin{equation}
\frac{{R_{per}^{crit}}}{{a_{in}}} = 2.8 \left( \frac{{(1 + q_{out})(1 + e_{out})}}{{(1 - e_{out})^{1/2}} }\right)^{2/5},
\end{equation} 
  where  $q_{out}= {m_{2}}/{(m_{0} + m_{1})}$
 is the outer mass ratio.  This equation gives an upper limit, which represents coplanar systems. Numerical simulations by \citet{mar99} show that non-coplanar systems tend to be more stable, and obtain from their simulations a reduction factor $f = 1 - 0.3 i/\pi$ (where $i$ is the relative inclination between the inner and outer orbit) to account for the increased stability of inclined orbits. \\ 
  This stability criterion can also be applied to the two planets around the primary star in HD188753 to place a lower limit on the distance of periastron of the second planet. 
 One can again substitute for $e_{out}$ the value of $e_{max}$, obtaining a lower limit on the semi-major axis of the second planet. This gives
\begin{equation}
   a_{2, \rm min}  =  2.8\times a_{1}\, \frac{ (\, (1 + q_{out}) (1 + e_{max})\, )^{{2}/{5}}\, }{ \,( 1 - e_{max})^{{6}/{5}}}.
\end{equation}\\
   The MA01 criterion is weakly dependent on the mass of the outer planet  $m_{2}$ through the ratio $q_{out}= {m_{2}}/{(m_{0} + m_{1})}$ in equation (18).  Therefore, the difference in the estimated value of  $a_{2, \rm min}$ obtained when using the two different values of  $m_{2}$ ($M_{J}; M_{\oplus}$) is very small and therefore is neglected.

\subsection{The Predicted Stability Region}
In Figure 3 we show  the analytically predicted stability region for a double-planet  system  with $m_{1} = m_{2} =  M_{J}$ in HD188753 where a hot Jupiter at $0.05$ AU represents the inner planet.  The upper limit ($a_{2, \rm max}$) is constrained by the \citet{hol99} criteria from which a value of  $\sim 1.23$ AU was obtained in the coplanar case.  
\begin{figure}
\epsscale{0.85}
\plotone{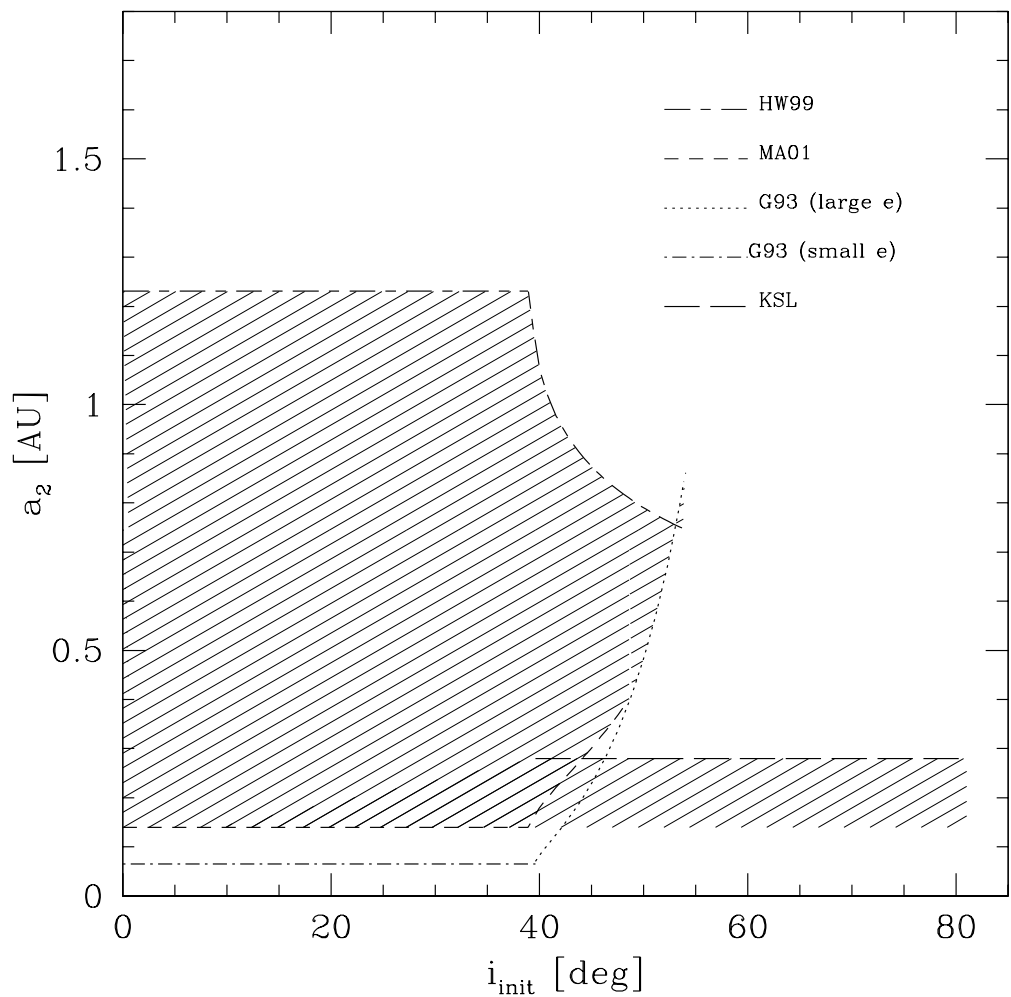}
\caption{The analytically predicted stability region for a second planet  in HD188753 with a hot Jupiter at $0.05$ AU.  The shaded region represents the most stringent limits expected for a second planet of mass $m_{2} = M_{J}$. The upper limit is constrained by the \citet{hol99} criteria (short dashed long dashed). The lower limit is constrained by the \citet{mar01} (dashed), and the \citet{gla93} (dotted (large e) and dotted dashed (small e)) criteria. The Kozai Suppression limit is shown in long dashed. }
\end{figure} 
For the lower limit on $a_{2}$,  both the
G93 and MA01 stability criterion were applied to the central star in HD188753 with two planetary orbits. The maximum eccentricity of the outer planet was calculated from  equation (1) for every value of the initial inclination $i_{init}$.  For the range $i_{init} < 40^{\circ}$, the value $e_{max} = 0$ was adopted.   Therefore, this range represents the expected stability criterion for initially circular orbits. 
The minimum semi-major axis allowed for the second planet derived from G93 is $a_{2,\rm min}\approx 0.065$ AU, while the MA01 sets a more stringent limit of $a_{2,\rm min}\approx   0.14$ AU. Applying the reduction factor ($f = 1 - 0.3 i/ \pi$) for  $i = 70^{\circ}$, reduces this value to $\sim 0.12$ AU.
Figure 3 also shows the Kozai Suppression Limit (KSL).  The  shaded area below this line is  the \emph{predicted Kozai-stable zone}  within which a second planet is expected to survive Kozai oscillations even in the presence of high relative inclinations ($\emph{i}_{initial} \geq 40^\circ$).

\section{Numerical Methods}

  The dynamics of a double-planet system in a tight binary with separation in the range $a_{b} = 12 - 20$ AU is studied.  We look into the effect a second planet may have on the orbit of a hot Jupiter in the presence of a close-in binary companion. We also investigate the stability criteria of this class of systems  for two different binary configurations.  The first, taken as a representative of the tightest binary in this class, resembles the stellar system HD188753 with separation $12$ AU and eccentricity $0.5$. The second is a $20$ AU equal-mass binary ($m_{0} =m_{b}=  M_\odot$) with zero eccentricity.   In all our integrated systems a hot Jupiter of mass $m_{1}=M_{J}$ is placed on an initially circular orbit at $0.05$ AU.  We consider two cases for a second planet:  a Jupiter mass planet (equal mass planets), and an Earth mass planet. We examine the possibility of a habitable planet in  HD188753  in the light of these results.  \\
   For integrations resembling HD188753,  the semi-major axis of the second planet was taken in the range  $0.055$ AU $< a_{2} <$ $1.8$ AU in bins of $0.1$ or $0.2$ AU in all regions excluding the closest to the inner orbit where we examine numerically the cases $a_{2} = 0.055, 0.06, 0.063, 0.066$ and $0.1$ AU. In the case of the  $20$ AU binary,  $a_{2} $ was varied between the values   $0.06$ AU and   $6.0$ AU in bins of $1$ or $0.5$ AU. 
     In all of our simulations, the initial planetary orbits are circular and coplanar. The common planetary orbital plane is chosen initially to make an angle with the binary orbit $ \emph i_{init}$, which was varied between $0^\circ$ and $70^\circ$ in steps of $10^\circ$ in most cases. 
     
We perform a total of $5460$ simulations representing $273$ different initial configurations. For each configuration, many realizations were constructed with different initial relative phases between the orbits, which were chosen randomly. The number of realizations varied between $5$ and $100$ depending on the configuration in hand.\\
     Integrations were performed using a modified version of the integration package MERCURY6.2 (\citet{cha99}). The code was modified to include GR effects as an additional force. We adopted for this force the standard textbook $\frac{1 }{ r^{4}}$ term  \citep[e.g,][eq. 25.16]{mis73}.
 Since our integrations are expected to involve close encounters, we use for all of our integrations the Bulirsch-Stoer integrator.  We monitor conservation of total energy and angular momentum for the duration of the integration. Energy degeneration was always maintained to $< 10^{-5}$ and fractional angular momentum change was always smaller than $< 10^{-7}$, excluding runs performed with the second planet's orbit close to the inner orbit ($a_{2}\leq 0.1$ AU).  In these cases, the fractional angular momentum change would fall in the range $10^{-6} - 10^{-5}$. 

 Since direct integrations of such tight systems involve many close encounters, and are computationally expensive, and since we are not interested in following the dynamical evolution of these systems after they become unstable, we optimize our computations by stopping them as soon as instability develops. We test the timescale for the onset of instability by integrating a system just outside the expected stability region. The system chosen for this test is one with  a Jupiter mass planet placed on an orbit with semi-major axis $ a_{2} = 1.4$ AU and a relative initial inclination of $ 40^\circ$. We show the instability timescale histogram for this system in Figure 4, which includes 363 different realizations. We find that
 $ 99 \% $ of these simulations hit instability by $6.6 \times 10^{5}$ years.
 Based on these results we chose for our integrations the integration time  $6.6 \times 10^{5}$ years to decide stability for a Jupiter mass planet.  This timescale corresponds to more than $10^4$ 
 binary cycles.  We also integrate systems with a Jupiter mass planet for as long as 6.6 Myr when this planet is near the Kozai suppression limit, since these systems tend to survive longer than their other counterparts.
 
   \begin{figure}
 \epsscale{0.85}
\plotone{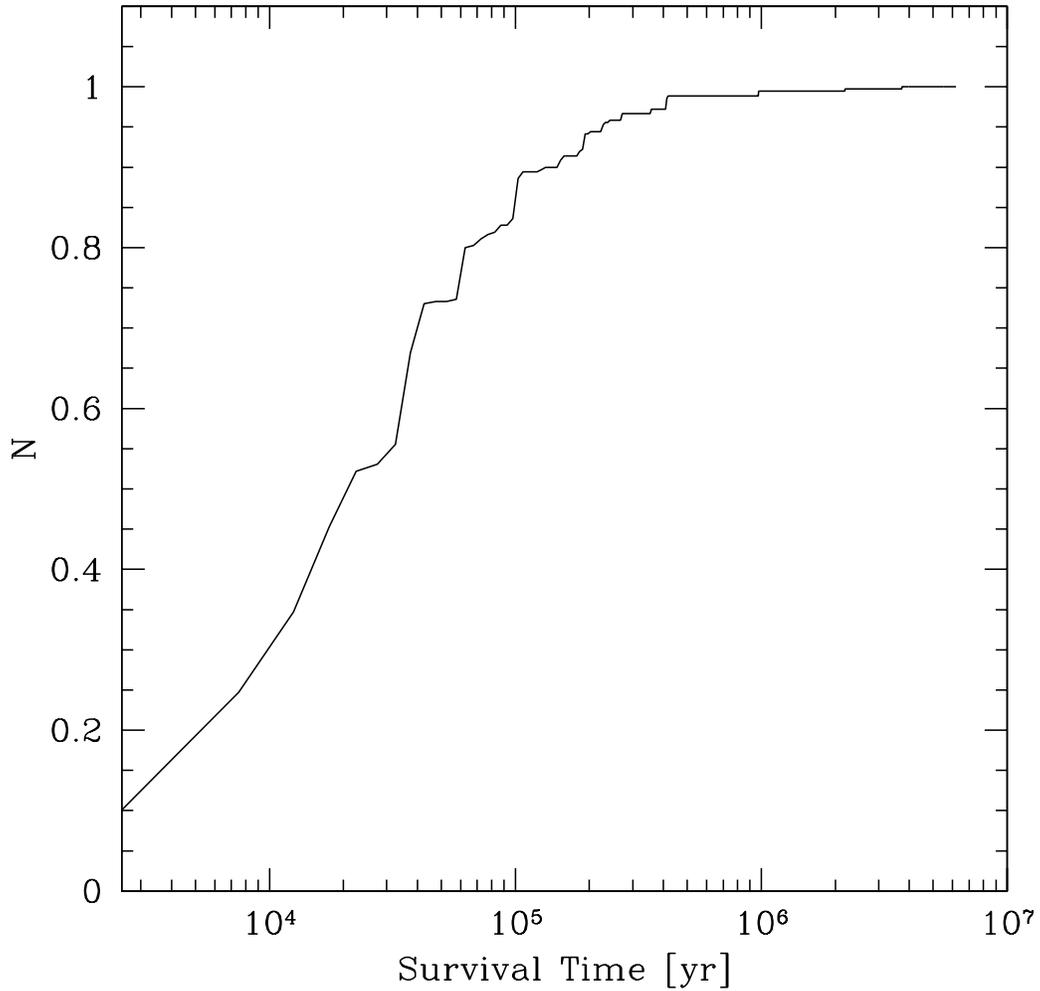}
\caption{ Instability timescale histogram for a Jupiter mass planet just outside the expected stability region, with $ a_{2} = 1.4$  AU and $\emph{i}_{init} = 40^\circ$ in a binary resembling HD188753.   "N"  is normalized to one and represents the cumulative fraction of systems that have already become unstable. We find that $ 99\% $ of systems which end up unstable hit instability before $6.6 \times 10^{5}$ yr. } 
\end{figure}

\section{Results and Discussion}
We test the stability region within a tight binary for a double-planet system composed of a Jupiter mass planet at $0.05$ AU and a second planet on a wider orbit. Two binary separations are investigated, $12$ AU and $20$ AU.   The stability region is tested for a Jupiter mass planet in the outer orbit for both binary separations.  In the case of  $12$ AU separation, the stability region for a potentially habitable  planet is also tested where $m_{2}=M_{\oplus}$. 

\subsection{ A $12$ AU Binary: The Case of HD188753}

 \subsubsection{Equal-Mass Planets}
 The stability region of the double-planet system described above in a binary system resembling HD188753 is investigated. 
Figure 5 summarizes  the results of all our simulations of this system spanning the whole expected region of stability for a second planet of Jupiter mass. It shows that as long as the orbits are highly inclined ($\gtrsim 40^\circ$), Kozai oscillations will always lead to instabilities in this system.
\begin{figure}
\epsscale{0.85}
\plotone{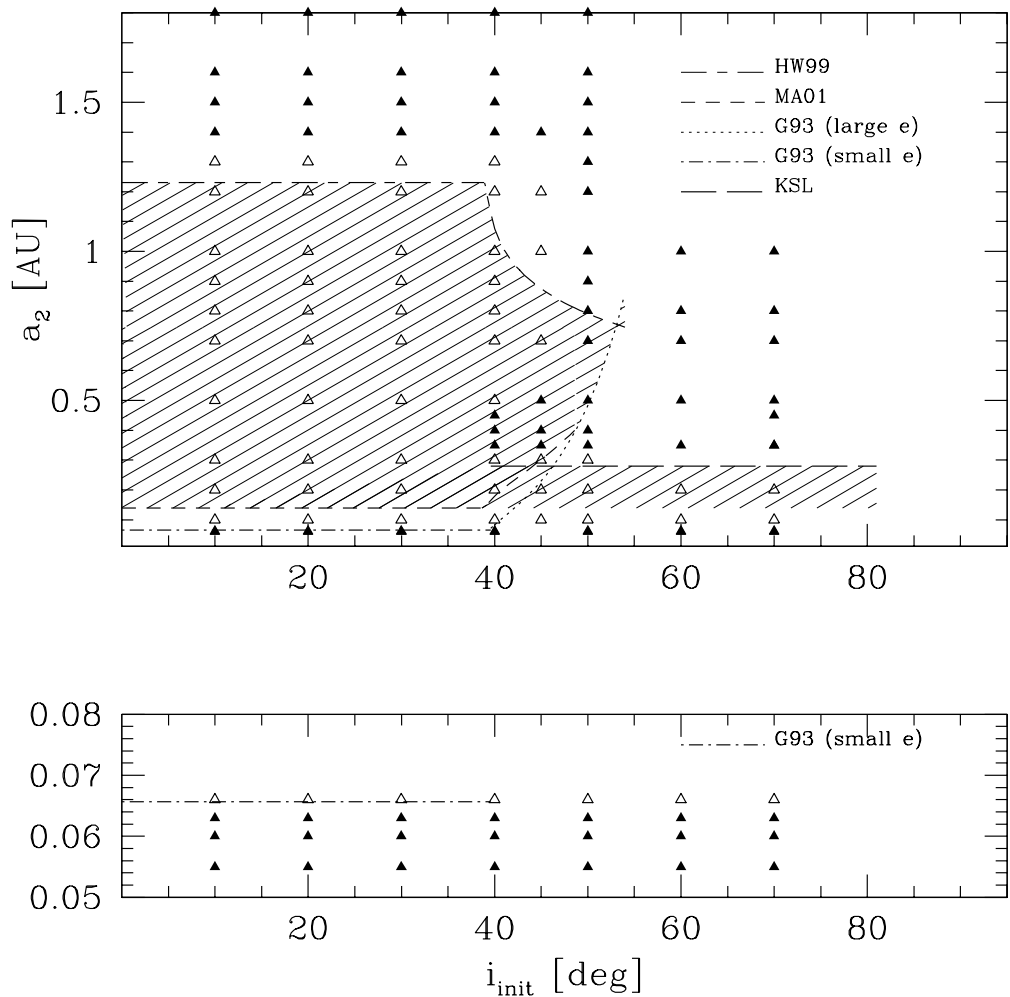}
\caption{The stability region for a second planet in HD188753 of mass $m_{2}=m_{1} = M_{J}$. Empty triangles represent stable systems while filled triangles represent systems which experience instabilities. All systems with the binary orbital plane highly inclined relative to the planetary plane are unstable, except for systems within the Kozai-stable zone.}
\end{figure}
    For small relative inclinations $i_{init} \leq 40^\circ$, the simulations give a \emph{lower limit} for the semi-major axis of this planet in the range $0.063$ AU $< a_{2, \rm min} < 0.066$ AU, which is consistent with the G93 predicted value of $0.065$ AU. 
The \emph{upper limit} lies in the range $1.3$ AU $ < a_{2, \rm max} < 1.4$ AU for the coplanar case, larger than the value predicted by HW99.  These results are obtained using full 4-body simulations which include planetary interactions with all other particles in the system and therefore, are expected to set more realistic limits on stability boundaries than do studies that treat planets as massless point particles. 

 Our results also confirm the existence of a \emph{Kozai-stable zone} within which stability is maintained even in the presence of large relative inclinations. This is seen in the region  $a_{2} = 0.1 - 0.3$ AU.  At $a_{2} = 0.3$ AU, we can see a transitional region between stability and instability as the system approaches the edge of the Kozai-stable zone.
We compare in Figure 6 the time evolution of both planets in two different cases.  
\begin{figure}
\epsscale{0.85}
\plotone{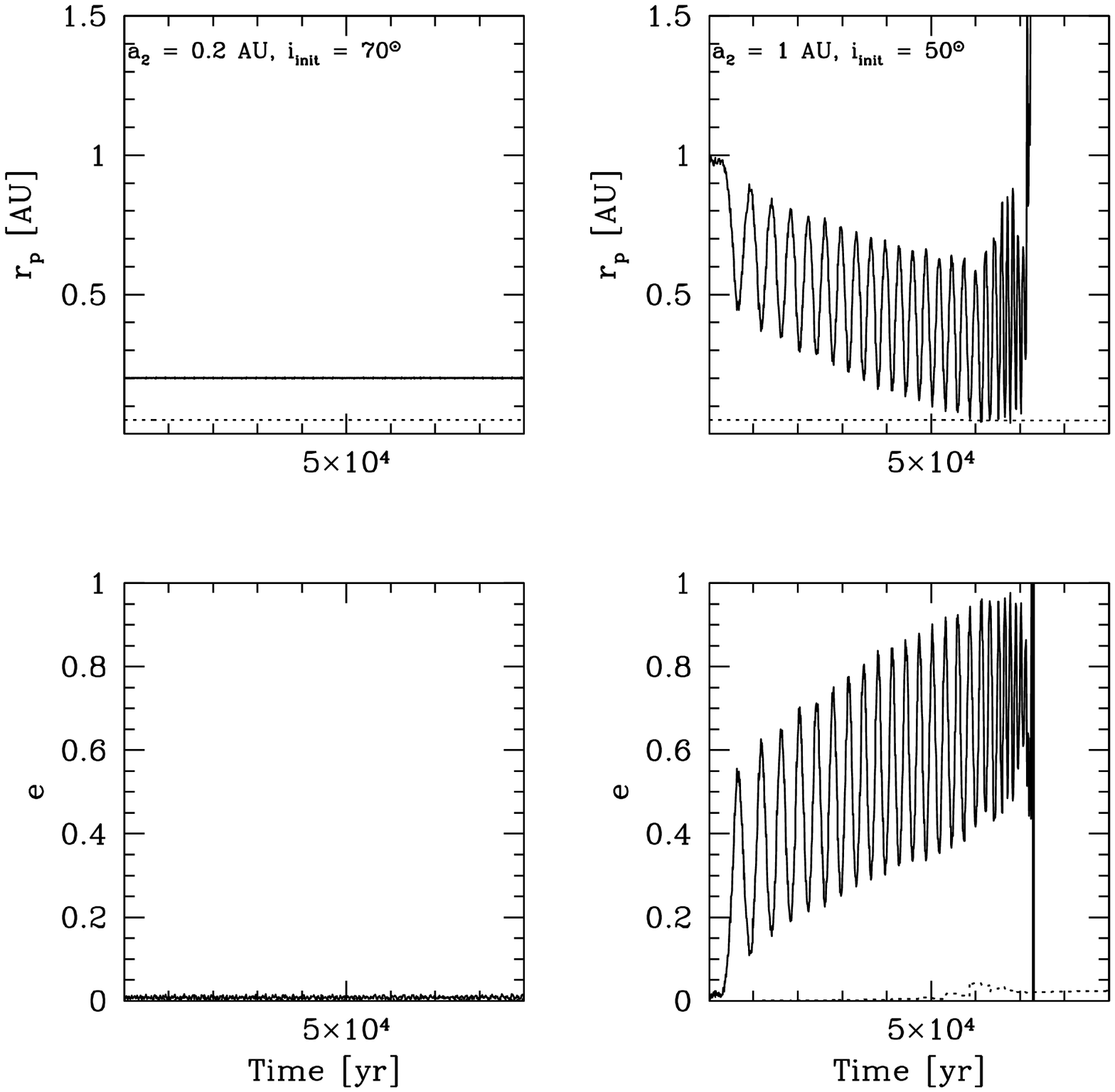}
\caption{ Orbital elements of both planets (inner (dotted) and outer (solid)) as they evolve with time for the following systems: - Left panel:  ($m_{1} = m_{2}= M_{J}$),  $a_{1} = 0.05 $ AU, $a_{2} = 0.2 $ AU (within the \emph{Kozai-stable zone}), and $i_{init}  = 70^{\circ}$. - Right panel: Same masses,  $a_{2} = 1.0$ AU , and $i_{init}  = 50^{\circ}$. The figure shows the distance at pericenter  ($ r_{p}$) and the eccentricity ($e$) of the planets. }
\end{figure} 
In the first the outer planet is within the Kozai-stable zone at $a_{2} = 0.2 $ AU. Although its orbit is highly inclined relative to the orbit of the companion, we see stable orbits with no sign of growth. The second case represents the outer planet outside the Kozai-stable zone at $a_{2} = 1.0 $ AU , where mutual gravitational perturbations become too weak to suppress Kozai oscillations. This case leads to the ejection of the planet within less than $80,000$ years. 

Figure 5 also shows a region of the explored space  ($0.35 < a_{2} < 0.8$ AU, and $i_{ini} \geq 40^{\circ}$) within which the planets are weakly coupled and therefore Kozai oscillations are induced in the inner orbit by interaction with the outer planet.  Large eccentricities are gained during these oscillations which eventually bring the inner planet to the surface of the star.  We discuss this effect in detail in \S 5.3.
 
\subsubsection{A Terrestrial Planet}

  There has been great interest lately in the question of habitability of exoplanets \citep[e.g.,][]{dav03,jia05,fat06,hag06,hag07}.  Although observations of Earth mass planets within the Habitable Zones (HZ) of Sun-like stars may not become reality in the very near future, numerical simulations to understand the dynamics and stability limits of these planets in different environments is necessary to guide future efforts. 
  
 It has been shown that  Earth-like planets can survive the migration of  a giant planet  \citep{man07, raym06}. Therefore, we look into the stability of  a double-planet system with a habitable planet  and a hot Jupiter in HD188753.   The central star in this system with mass $1.06$  $M_{\odot}$ is of G-type with a habitable zone (HZ) between $0.95$ AU and $1.37$ AU \citep{kas93}. Therefore, our predicted stability region (Fig. 5) does leave space for a habitable planet in this system, but only for small values of the initial relative inclinations of the orbits.
 
 Figure 7 shows the results of our simulations performed for a second planet of mass $m_{2}=M_{\oplus}$ in HD188753.  Since an Earth mass planet is three orders of magnitude less massive than a Jupiter mass, perturbations from the companion are expected to become less dominant leaving more space for stability. Therefore, a maximum integration time of  $6.6 \times 10^{6}$ yr was used for all systems shown to account for any delayed instabilities. 
 \begin{figure}
\epsscale{0.85}
\plotone{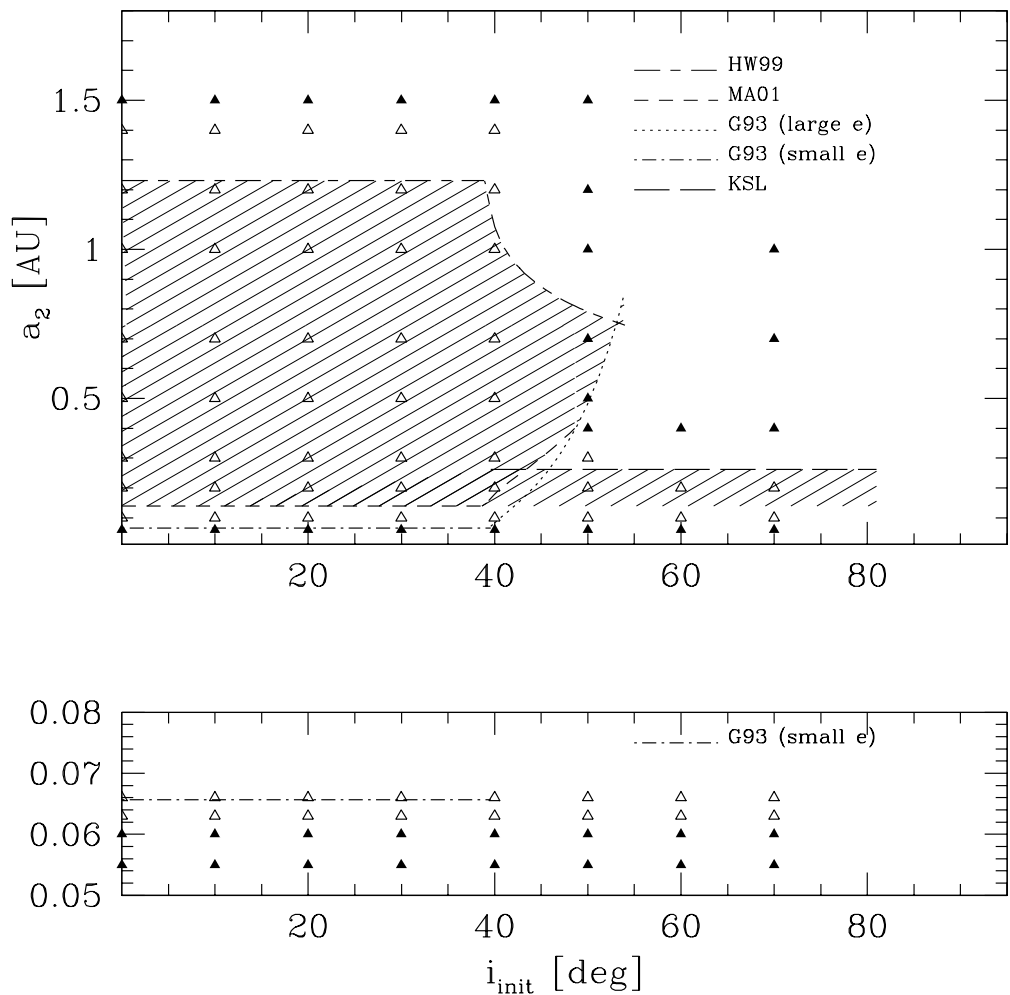}
\caption{The results of all numerical simulations performed for a system similar to HD188753 with a second planet of mass $ = 1 M_{\oplus}$ . Empty triangles represent stable systems while filled triangles represent unstable systems. The stable region includes the HZ of this star, but only for low relative inclinations among the orbits. }
\end{figure}
Our results show that the upper limit for stability of an Earth mass planet at low relative inclinations ($i_{init} \leq 40^\circ$) is larger than that for a Jupiter mass planet  and lies in the range $1.4$ AU $ < (a_{2})_{max} < 1.5$ AU, confirming the dependence of perturbations, and therefore stability limits,  on the planetary-masses. The HZ of the central star in HD188753 lies completely within the stable region at low inclinations.  On the other hand, the \emph{Kozai-stable zone} is located outside the HZ, leaving no space for a habitable planet when high relative inclinations are present. 
 At the lower limit of $a_{2}$,   although an Earth mass planet can survive down to a region between $0.06$ AU and $0.063$ AU such a planet will be extremely hot and inhabitable. 
 
 In addition to dynamical stability, an important condition for habitability is climate stability which can be comprimised if the planet's orbit becomes eccentric.  This has the potential to drive the planet's orbit outside the HZ at apastron or/and periastron, or may lead to extreme heat variations on the surface of the planet. 
  We show in Figure 8 the orbital elements of an Earth-mass planet in this system with low relative inclination and initially on a circular orbit at $a_{2} \sim 1.0 $ AU. It experiences mild 
  eccentricity oscillations with an amplitude of $e \sim 0.1$, with no sign of growth.  
\begin{figure}
\epsscale{0.85}
\plotone{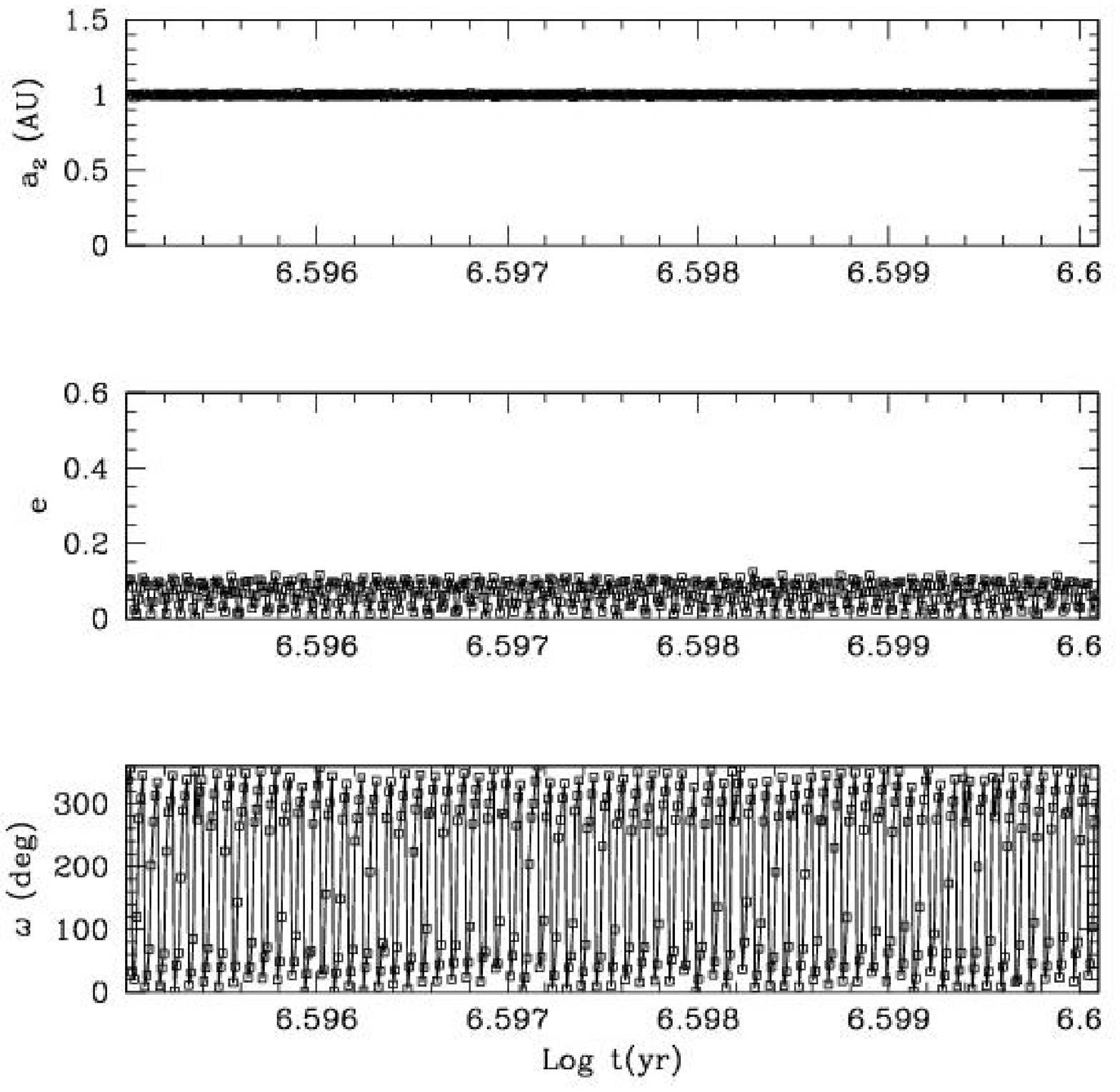}
\caption{Orbital elements of the outer planet in a double planet system in HD188753 with the outer planet in the Habitable Zone of the primary. System elements are : $a_{1}=0.05$ AU; $a_{2}=1.0$ AU; $m_{1}=M_{J}$;  $m_{2} =M_{\oplus}$ and $i_{init}  = 0^{\circ}$ . The figure shows from top to bottom:  semi-major axis, eccentricity and argument of pericenter of the outer planet. }
\end{figure}
Still, habitability  of this planet can become questionable since such an eccentricity can drive the planet near the edges of the HZ at periastron and apastron.  Therefore, the presence of  a perturber to the Keplerian motion of a potentially habitable planet can alter the positions of the outer and the inner edges of the HZ through their effect on the orbit.  In this case, the outer edge of the HZ is expected to move inward to  $ a_{2} \sim 1.24$ AU from the primary while the inner edge move from $a_{2}\sim 0.95$ AU to $\sim 1.0$ AU, altering the width of the HZ by $\sim 40\%$.

 \subsection{ A $20$ AU Equal mass Binary}
    It has been shown by \citet{jan07} that a stellar system as tight as HD188753  is unlikely to support a disk massive enough to form a Jovian planet \emph{in situ}. Therefore,  the presence of a planet in such a system would be indicative of  a dynamical history.  On the contrary, the truncated disk around a binary with slightly larger separation such as $\gamma$ Cephei (separation $20$ AU and eccentricity $0.4$) has been shown  \citep{jan08} to have sufficient  mass to form such a planet.  
      Such wider binaries may host giant planets and still have no active dynamical histories, and therefore, maintain  dynamical stability with multiple planets which are coplanar with the binary orbit.  Therefore, a  binary with separation $20$ AU, represents an upper limit in this study to a class of binaries with  low separation in which the presence of a hot Jupiter is questionable.  
      
         We show in this section, our results for the stability region of a double planet system in a binary with two stars of solar mass orbiting at $20$ AU.  The planetary system used in this test is the same system used in \S5.1.1 in which two coplanar, equal mass planets of Jupiter size are place on circular orbits.  The inner planet is a hot Jupiter at $0.05$ AU, while the semi-major axis of the second planet, $a_{2}$, is varied along the expected range. 
 The stability region is also tested here in the two dimensional space [$i_{init}, a_{2}$] where $i_{init}$ is the initial relative inclination between the binary orbit and the common planetary plane.  We apply the same criteria as in \S 3 to predict the stability region for the outer planet when the inner orbit is fixed. These results are shown in Figure 9 where one can see that the stability region for this system is similar to that of HD188753 in its structure but varies in scale.  The larger binary separation has allowed more space for the outer planet to survive.  In some cases the outer planet can experience Kozai oscillations (with moderate eccentricities) for as long as $1$ Myr without disrupting the system as long as it does not reach the upper limit for stability at apastron. 
 The Kozai stable zone expands with the binary separation in general since mutual planetary interactions become more dominant relative to the Kozai effect.  In this system the edge of the Kozai stable zone is at $a_{2} = 0.5$ AU.  
 We also see that the two planets are placed in the "weakly coupled" dynamical class in a small region of the shown parameter space  ($ 0.5 \gtrsim a_{2} \lesssim 0.8$ AU and $i_{init} = 50^\circ$) within which the inner planet experiences Kozai oscillations induced by the outer planet. The width of this region is narrower than that observed in the case $a_{b} = 12$ AU  (see \S 5.3 for a more detailed discussion). 
  \begin{figure}
\epsscale{0.85}
\plotone{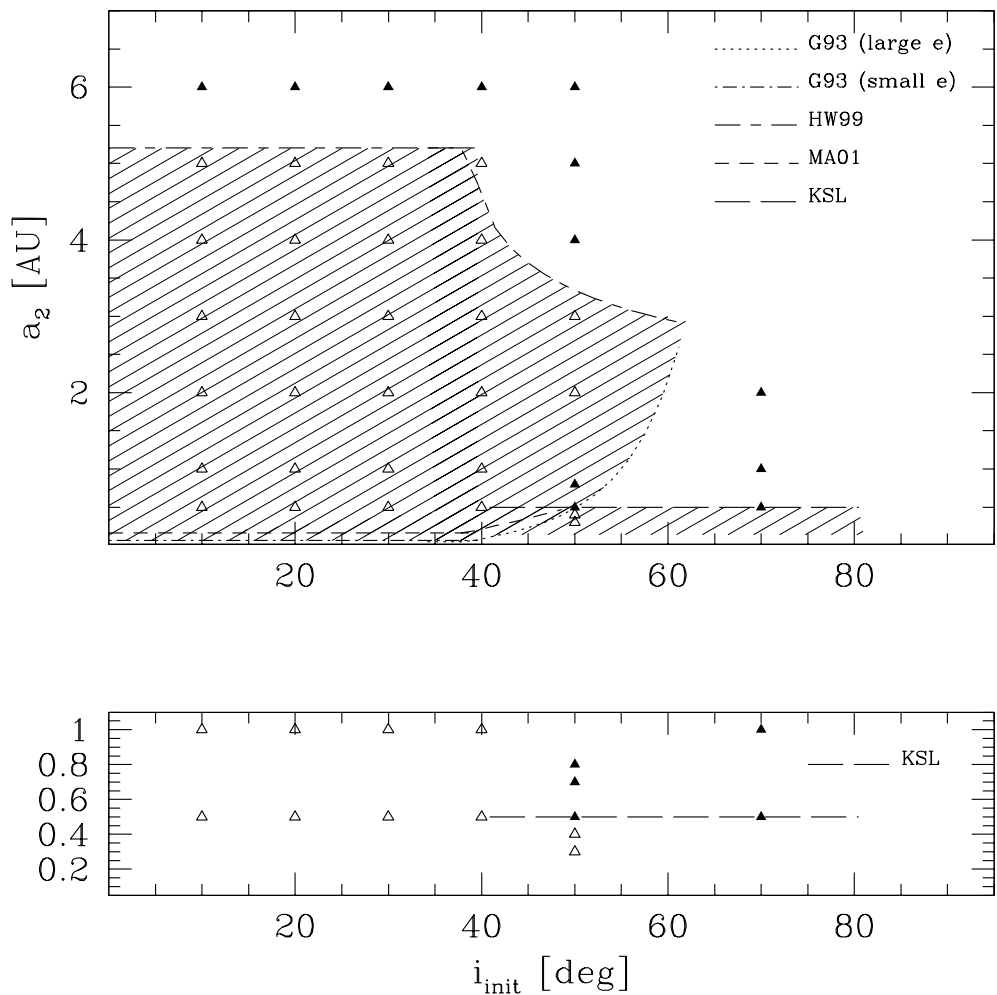}
\caption{The stability region for a double planet system in a binary with parameters: Two stars of Solar mass, $a_{b} = 20$ AU, and $e_{b} =0$.   Planetary masses are $m_{2}=m_{1} = M_{J}$ initially on circular coplanar orbits. The inner planet is a hot Jupiter at $a_{2} = 0.05$ AU. Empty triangles represent stable systems, while filled triangles represent the unstable systems.}
\end{figure}

\subsection{Induced Eccentricity Oscillations in a hot Jupiter}
 
  We test the possibility of inducing Kozai oscillations in the hot Jupiter orbit in our systems by propagating perturbations from the stellar companion through a second planet of equal mass.  We find that  inducing large eccentricities in the hot Jupiter orbit is more likely in binaries with smaller separation.  In the case of a $12$ AU binary separation this effect is observed within  
the region of the explored phase space defined by  ($0.35 \leq a_{2} \geq 0.8$ AU, and $i_{ini} \geq 40^{\circ}$). 
This behavior is not observed within the  \emph{Kozai-stable zone}   ($a_{2} < 0.3$) AU, and shows its strongest behavior in the region  ($0.4$ AU $\leq a_{2} \leq 0.5$ AU).  As the distance between the planets increases, this effect gets weaker until it disappears beyond $a_{2} > 0.8$ AU.   Figure 10 shows the time evolution of the orbital elements of a system representative of this behavior with the outer planet at $a_{2}=0.5$ AU and $i_{init}=50^\circ$.   
  \begin{figure}
\epsscale{0.85}
\plotone{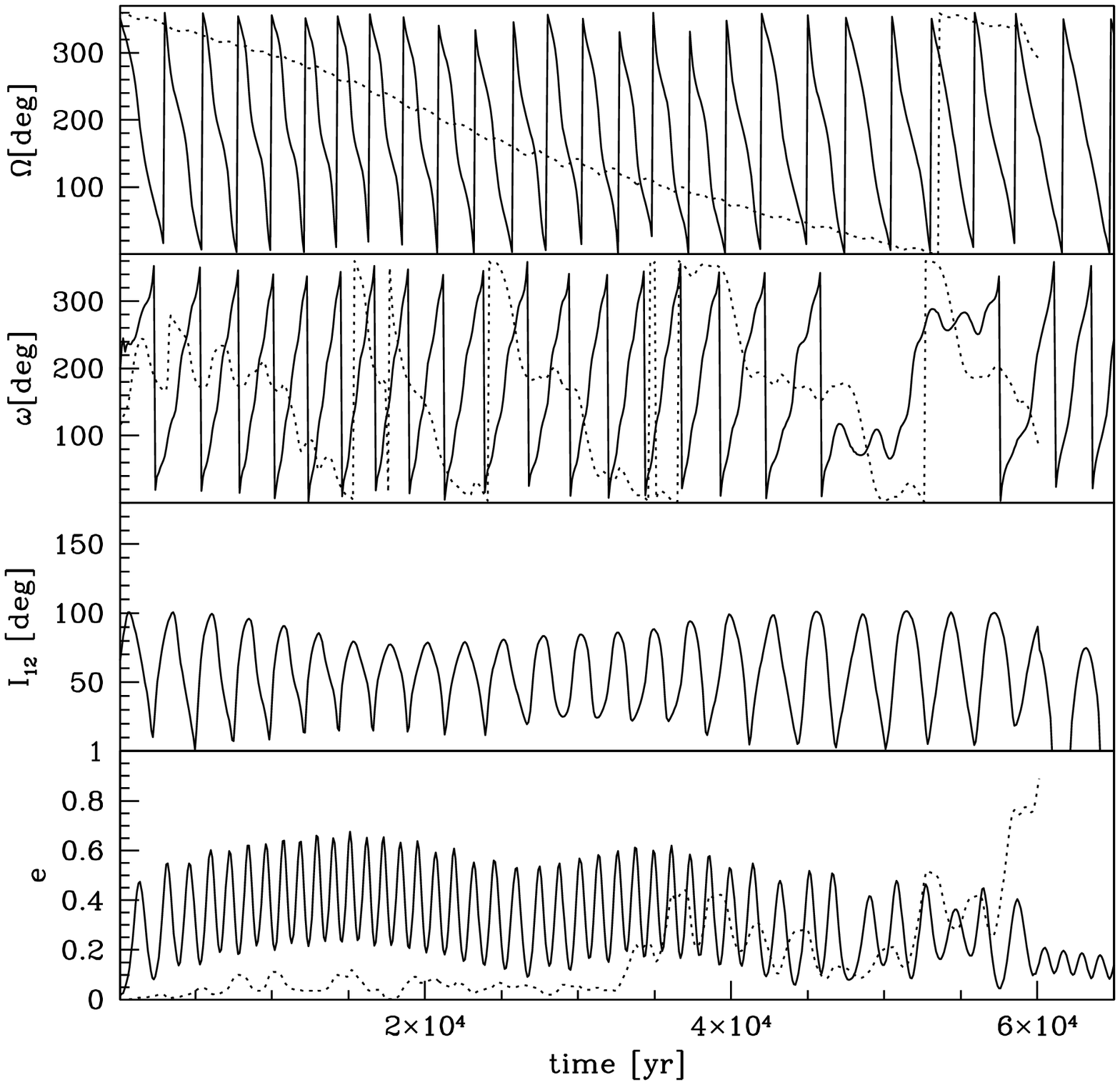}
\caption{A case representative of the systems which experience induced eccentricity in the inner orbit, showing orbital elements of both planets, inner (dotted line) and outer (solid line). The initial system parameters are: $m_{1} = m_{2}= M_{J}$,  $a_{1} = 0.05 $ AU, $a_{2} = 0.5 $ AU, $e_{1}=e_{2}=0$, and $i_{init}=50^\circ$. The outer planet is experiencing Kozai oscillations induced by the stellar companion. Due to fast nodal precession of the outer planet, the mutual inclination between the planets $I_{12}$ grows, and oscillates with the same period as $\Omega_{2}$.   The amplitude of these oscillations grows to a value twice the common initial inclination relative to the binary plane,  $i_{init}$, causing the inner planet's eccentricity to grow chaotically. }
\end{figure}
The outer planet is experiencing Kozai oscillations induced by the stellar companion. The nodal precession of this planet creates a split in the planetary orbits and the mutual inclination between the planets, $I_{12}$, grows and oscillates with the same period as $\Omega_{2}$.   The amplitude of these oscillations is twice the common initial inclination relative to the binary plane $i_{init}$.  Angular momentum exchanged between the planetary orbits, propagating the angular momentum transfer from the binary leads to mutual planetary Kozai oscillations.  Since the angular momentum of the perturber ($m_{2}$ in this case) is changing, the eccentricity oscillations induced in the inner orbit lead to chaotic growth.  
As the planet approaches the central star, it should experience strong tidal dissipation  and may be circularized into a tighter orbit. Our simulations do not include tidal effects and therefore, we cannot predict the final outcome of these events.     
In general,  tidal torques are not significant in the evolution of a circular orbit at $\sim 0.05$ AU where GR is the most dominant force.  Once a planet gains significant eccentricities driving it very close to the surface of the star at periastron, tidal forces may become a dominant factor but do act on longer timescales when compared to Kozai precession timescales. 

  To demonstrate the interplay between the different forces acting on the inner planet, we show in Figure 11
 \begin{figure}
\epsscale{0.85}
\plotone{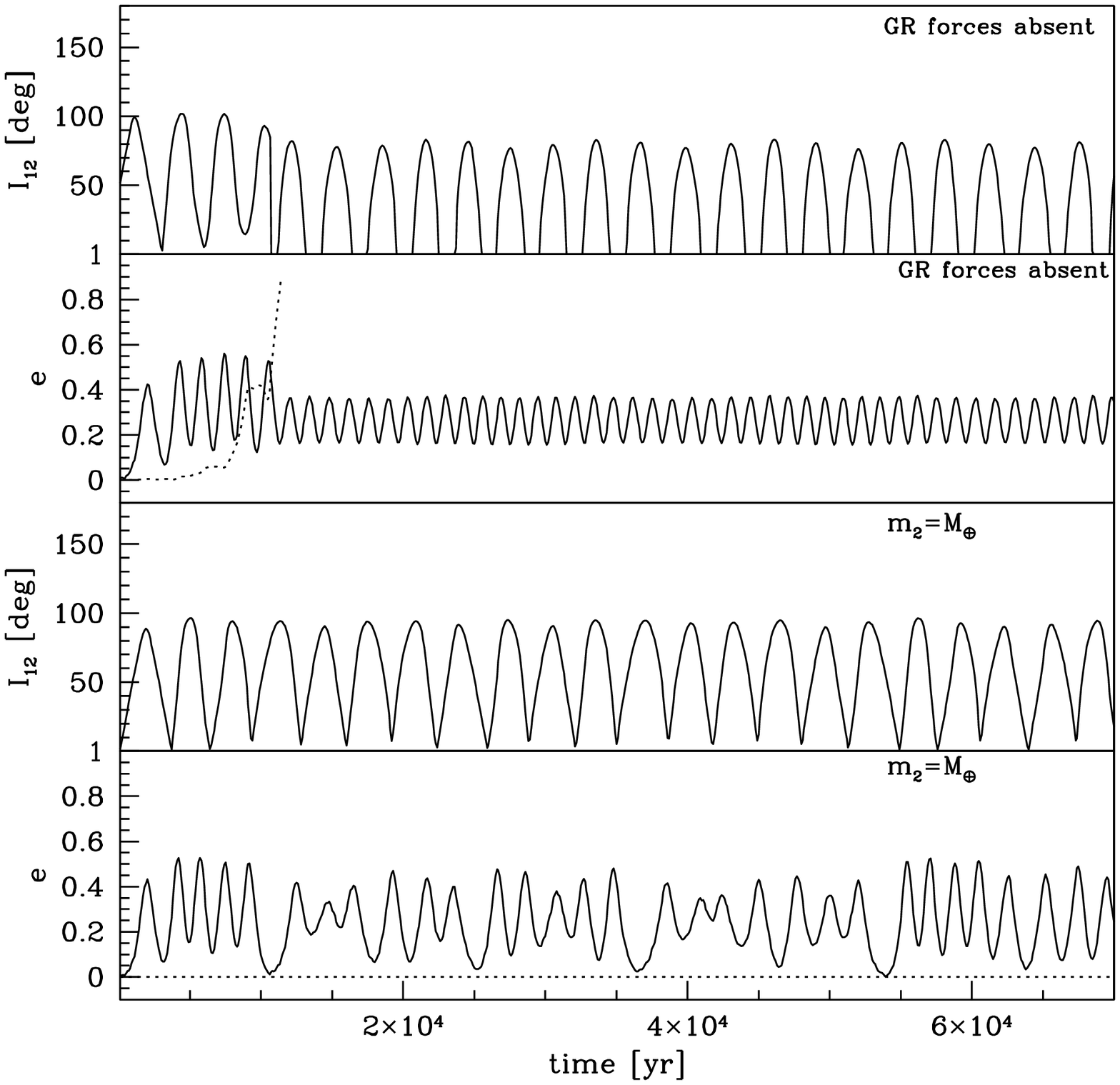}
\caption{ Eccentricity of both planets (outer in (solid) and inner in (dotted)) and mutual inclination between planets $I_{12}$ shown for the same system in Figure 10, but with GR forces absent in the top two panels and and $m_{2}=M_{\oplus}$ in the lower two panels.  } 
\end{figure}
 the relative inclination between the planetary orbits and the eccentricities of both planets in a system similar to that shown in Figure 10, but once with the GR forces removed from the calculation (top two panels) and second using an Earth-mass planet in the outer orbit (lower two panels).  The former shows the faster growth of eccentricity when compared to Figure 10 and therefore demonstrates the goal played by GR forces in stabilizing the inner orbit against other perturbations, while the later confirms the dependency of this mutual planetary Kozai effect on the planetary masses.  
 
 Inducing such oscillations into the orbit of a hot Jupiter requires specific initial conditions to be satisfied (see \S 2.4).  We find that the region in parameter space that allows these conditions to be satisfied is narrower for wider binaries. We show in Figure 12 the range in semi-major axis of the outer planet that will allow induced Kozai oscillations in the inner orbit as a function of binary separation.   This is shown for an inner planet on an initially circular orbit at $a_{1}=0.05$ AU, and a binary orbit with inclination $i_{init} = 50^{\circ}$ relative to the common planetary plane. All binaries in this figure share the following parameters ($m_{0} = m_{b}= M_{\odot}, m_{1}=m_{2}=M_{J}$ , and $e_{b} = 0$).
 \begin{figure}
\epsscale{0.85}
\plotone{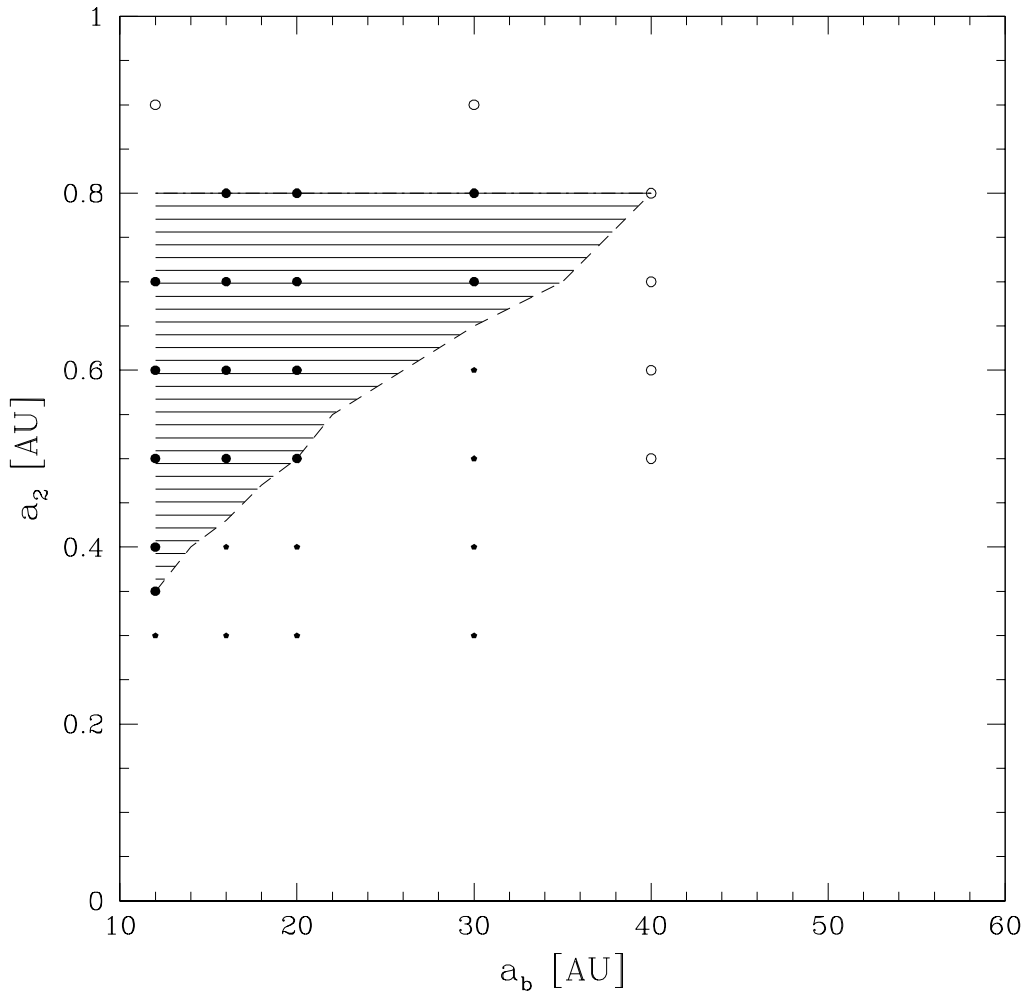}
\caption{ The region in parameter space which allows Kozai oscillations to be induced in a hot Jupiter orbiting at $0.05$ AU.  On the x-axis, $a_{b}$ represents the variable binary separation. Binary parameters for all binaries are:  $m_{0} = m_{b}= M_{\odot}, m_{1}=m_{2}=M_{J},  e_{b} = 0$, and $i_{init} = 50^{\circ}$.  The variable semi-major axis of the outer planet $a_{2} $ is plotted on the y-axis.  In order to overcome GR precession and induce Kozai oscillations in the inner orbit, the outer planet must lie below the line $a_{2} = 0.8$ AU. The dashed line represents the limit of the \emph{ Kozai-stable zone} as a function of binary separation.  Inside the dashed region between the two limits,  both planets experience Kozai oscillations (outer planet induced by the binary and the inner planet induced by the outer planet). All circles represent results of numerical simulations.  The large filled circles are systems in which the outer planet experiences Kozai oscillations and always lead to eccentric hot Jupiters. Small filled circles are systems in which the outer planet is shielded from the  Kozai effect, but in  less than $50\%$ of the simulations we observe induced eccentricities to the hot Jupiter.  Empty circles represent systems which did not show any sign of this effect.} 
\end{figure}
These results show that tighter binaries are more likely to produce this effect and that  a cut-off binary separation $a_{b, \rm cut}$ exists beyond which it is not observed.  For a hot Jupiter of mass $M_{J}$ and semi-major axis $a_{2} =0.05$ AU, this cut-off binary separation lies in the range $ 30 < a_{b, \rm cut} < 40$ AU. As the inner planet's orbit is made wider, mutual planetary interactions grow larger which should allow a wider binary to induce eccentricities through weak coupling between the planets into the inner orbit.

\section{Summary and Direction for Future Work}

 The presence of a giant planet in a compact stellar system with separation $\lesssim 20 $ AU is indicative of  an active dynamical history in which the binary companion is either captured or moved inward to form a harder binary during an encounter after the planet has already formed. Such an event is likely to result in high relative inclinations between the orbits of the planet and the companion and may lead to Kozai oscillations in the planetary orbit/s.  Since multi-planet systems seem to be common, it is likely that the primary in such a system may be harboring more than one planet at the time of its encounter with a stellar companion.  In such a scenario,  Kozai oscillations can easily lead to instabilities including collisional events and ejections.  Therefore, such encounters which may occur in dense stellar systems when planetary systems are in their infancy, can be very important events in shaping the final outcome of planetary systems and leading to the large diversity observed in their characteristics.  
 
 We investigate the dynamical evolution of a double-planet system in a class of tight binaries with separation in the range $12 - 20$ AU assuming relative inclinations between the common planetary plane and that of  the binary in the range $0 - 70^\circ$. Our simulations reveal the large parameter space involved with a double-planet system in a binary such that the final outcome of a system is highly sensitive to the initial conditions. The stability criteria are tested for two systems with binary separations $12$ and $20$ AU.   We find that if the orbit of the stellar companion is highly inclined ($\gtrsim 40^{\circ}$) relative to the common planetary plane, the system is likely to experience instabilities unless the outer planet is situated in a narrow \emph{ Kozai-stable zone} within which gravitational perturbations with the inner planet are strong enough to suppress the Kozai mechanism. At inclinations smaller than the critical Kozai angle, we find that a binary as tight as $12$ AU can remain stable and an Earth-like planet can survive within the HZ of the central star up to $6$ million years.  
 
  The \emph{eccentric hot Jupiter} phenomenon is a puzzling observation since hot Jupiters are expected to be the result of a migration process during which circularization of the orbit occurs.  We suggest that eccentricities can be pumped up into a hot Jupiter by inducing Kozai oscillations into its orbit through a weak coupling mechanism with a second planet in the presence of a close-in binary companion.   A fine tuning of the initial conditions can produce this effect allowing perturbations from the stellar companion to propagate through the planetary system into the inner orbit.  We find that tighter binaries are more likely to produce this effect, and that for a given hot Jupiter orbit, a cut-off binary separation can be found beyond which inducing Kozai oscillation into its orbit becomes impossible. 
The final outcome of such a scenario as the inner planet gains large eccentricities cannot be predicted accurately without the inclusion of tidal torques. These forces become important as the planet approaches the primary at periastron during cycles of large eccentricities.  Circularization into a tighter orbit is one possibility that seems unlikely if the outer planet remains in the system undergoing nodal precession and inducing oscillations in the inner orbit. This is since the timescales associated with these oscillations are  orders of magnitude shorter than tidal circularization timescales.
  If the outer planet reaches the edge of the Hill's sphere of the primary at apastron and is ejected from the system, the inner orbit may have enough time to experience tidal circularization during the life-time of the star. 
  Explaining eccentric hot Jupiters by such a mechanism requires the inclusion of tidal circularization effects in more comprehensive studies of binaries harboring double-planet systems. Such studies may require numerical testing over longer timescales.  This mechanism can also be used to explain the higher eccentricities among planets hosted by binary systems in general and the relative paucity of planets in binaries with near-circular orbits (for periods greater than $40$ days) \citep{egg04}. This is because the wider the orbit of  the inner planet the easier it is to induce eccentricities into it by weak coupling with another planet on a wider orbit.
  
        Tight binary systems with planets constitute a unique class of binaries which can allow us to test theoretical models of planetary formation and evolution. Therefore, more comprehensive studies including multiple planets and stellar collisions are necessary for a better understanding of the forces that were responsible for shaping planetary systems with a large diversity in characteristics. \\
     Finally, we can benefit greatly from better observational constraints on these systems, especially orbital inclinations. 

  The authors gratefully acknowledge the Dr. Ralph and Marian C. Falk Trust and the National Science Foundation Grant AST-0507727 for their support of this research.
We would like to thank Genya Takeda  for useful discussions and John Fregeau for reading the manuscript.

\end{document}